\documentclass[12pt,preprint]{aastex}
\newcommand{\kms}{{\,\rm km\,s}^{-1}}
\newcommand{\etal}{{\it et al.}\,}
\newcommand{\ksm}{{\,\rm km} {\rm~s}^{-1} {\rm~Mpc}^{-1}}

\def\la{\mathrel{\hbox{\rlap{\hbox{\lower4pt\hbox{$\sim$}}}\hbox{$<$}}}}
\def\sun{\hbox{$\odot$}}

\begin{document}

\title{Can Old Galaxies at High Redshifts and Baryon Acoustic Oscillations Constrain $H_0$?}
\author{J. A. S. Lima\footnote{limajas@astro.iag.usp.br},  J. F. Jesus\footnote{jfernando@astro.iag.usp.br}, J. V. Cunha\footnote{cunhajv@astro.iag.usp.br}}
\affil{Departamento de Astronomia, Universidade de S\~ao Paulo, USP,
\\ 05508-900 S\~ao Paulo, SP, Brazil}

\begin{abstract} 
A new age-redshift test is proposed in order to constrain $H_0$ with basis on the existence of old high redshift galaxies (OHRG). As should be expected, the estimates of $H_0$ based on the OHRG are heavily dependent on the cosmological description. In the flat concordance model ($\Lambda$CDM), for example, the value of $H_0$ depends on the mass density parameter $\Omega_M=1 - \Omega_{\Lambda}$.  Such a degeneracy can be broken trough a joint analysis involving the OHRG and baryon acoustic oscillation (BAO) signature. In the framework of the $\Lambda CDM$ model our joint analysis yields a value of $H_0=71\pm 0.04\kms$ Mpc$^{-1}$ ($1\sigma$) with the best fit density parameter  $\Omega_M=0.27 \pm 0.03$. Such results are in good agreement with independent studies from the {\it{Hubble Space Telescope}} key
project and the recent estimates of WMAP, thereby suggesting that
the combination of these two independent phenomena provides an
interesting method to constrain the Hubble constant.
\end{abstract}
\maketitle


\newpage

\section{INTRODUCTION}
\label{sec:intro}
The determination of $H_0$ has a practical and theoretical
importance to many astrophysical properties of galaxies, clusters and quasars,
and plays an important role for several cosmological calculations and others physical rulers, like SNe Ia, the present energy density scale, primordial nucleosynthesis, and the age of the Universe (Peebles 1993; Peacock 1999; Freedman 2000). However, Spergel {\it{et al.}} (2007) have shown that CMB studies can not supply strong constraints on the
value of $H_0$ on their own. This problem occurs due to the high degree of
degeneracy on the parameter space (Tegmark {\it{et al}} 2004),
and may be circumvented  only  by using independent measurements of
$H_0$ (Hu 2005, Cunha {\it et al.} 2007). In these circumstances, it is particularly relevant to obtain accurate and independent bounds on the value of $H_0$ from physics relying on many different kinds of observations. 

On the other hand, with the new optical and infrared techniques together the recent advent of large telescopes became  possible to estimate the ages of high redshift objects like galaxies and quasars. The subsequent discovery of some evolved high-z objects  (Dunlop et al. 1996, Spinrad et al. 1997; Bruzual \& Magris 1997; Dunlop 1999; Nolan et al. 2001) have been used by many authors to put limits on the cosmological parameters.  Alcaniz and Lima (1999, 2001), by using only two old (extremely red) high redshift galaxies (53W091 and 53W069) dated by Dunlop and collaborators  obtained $\Omega_{\Lambda} \geq 0.5$ for a flat $\Lambda$CDM model, a result more stringent than the ones derived from Type Ia Supernovae. They also derived limits on the cosmic equation of state parameter by considering a flat cosmology driven by cold dark matter and a X-matter dark energy component ($p=\omega\rho, \omega=constant$). As a conclusion, it was argued that if the ages of those objects were correctly estimated, the first formation era is pushed back to extremely high redshifts. Hasinger, Schartel and Komossa (2002) also reported  the discovery of the quasar APM 08729 + 5255 at $z=3.91$ with an estimated age of 2-3 Gyr. Later on, this result was independently confirmed by  Fria\c{c}a {\it et al.} (2005) by using a chemodynamical model for the evolution of spheroids. They shown that a minimal age of $2.1$ Gyr is set  by the condition that the $F_e/O$ abundance ratio of the model reaches $3.3$ (normalized to solar values) which is the best fit values obtained by Hasinger and coworkers. Further, several authors have also claimed that the quasar APM 08729 + 5255 is an extremely stringent object for any realistic cosmological model (Alcaniz, Lima \& Cunha 2003; Cunha \& Santos 2004, Deepak and Dev 2006). In this regard, it is of particular interest the sample of 20 old passive galaxies published by the team of the Gemini Deep Deep Survey (GDDS).  As noticed by McCarthy et al. (2004), such data are strongly suggesting  that the galaxies have been evolving passively since their initial burst of star formation. More recently, the GDDS sample has been used by Dantas {\it et al.} (2007) to put  constraints on the dark energy equation of state ($w$) through a joint analysis involving the inferred lookback time measurements from the ages of the galaxies plus the total expanding age of the Universe as obtained from CMB data.

In this letter, we derive new constraints on the Hubble constant $H_0$, by using  
the GDDS sample complemented by two additional data set recently released (Roche {\it et al.} 2006, Longhetti {\it et al.} 2007).
A statistical age test is performed to constrain the ($h, \Omega_m$) plane assuming that the Universe is well described by the cosmic concordance model ($\Lambda$CDM). Due to a degeneracy on the space parameter, more stringent constraints are obtained through a joint analysis involving the different OHRGs samples and the current SDSS measurements of the baryon acoustic peak (Eisenstein {\it{et al.}} 2005). In this concern, we recall that the baryon acoustic oscillations (BAO) method is \emph{independent of the Hubble constant} $H_0$ but it is heavily dependent on the value of $\Omega_M$, and, as such, it contributes indirectly to fix the value of $H_0$. Therefore, one can apply the BAO signature to break the degeneracy of the mass density parameter ($\Omega_M$). The BAO signature comes out because the cosmological perturbations excite sound waves in the relativistic plasma, thereby  producing the acoustic peaks in the early universe. Eisenstein
{\it{et al.}} presented the large scale correlation function
from the Sloan Digital Sky Survey (SDSS) showing clear evidence for
the baryon acoustic peak at $100 h^{-1}$ Mpc scale, which is in
excellent agreement with the WMAP prediction from the CMB data. As we shall see, both set of data (OHRG sample and BAO signature) provide an interesting method to  constrain the Hubble constant.

\section{Basic Equations and Sample}

Let us now consider that the Universe is described by a flat
Friedmann-Robertson-Walker (FRW) geometry ($c=1$)
\begin{equation}
 ds^2=dt^2-a^2(t)\left[dr^2+r^2(d\theta^2+sin^2\theta d\phi)\right],
\end{equation}
as theoretically predicted by inflation and required by the WMAP observations (Spergel {\it et al.} 2003; 2007).

In such a background, the age-redshift relation for a model driven by cold dark matter plus a cosmological constant ($\Lambda$CDM) has only two free parameters ($H_0,\Omega_{M}$). It 
can be written as 
\begin{eqnarray}
t(z;h,\Omega_M)& = &H_{0}^{-1}\int_{0}^{(1 + z)^{-1}}{dx
\over
x\sqrt{\Omega_{\rm{M}}x^{-3} + (1 - \Omega_{\rm{M}})}}, 
\end{eqnarray}
where $h=H_0/100$ km s$^{-1}$ Mpc$^{-1}$, and from now on the subscript $0$ denotes present day quantities.  Note that for
$\Omega_{\rm{M}} = 1$ the above expression reduces to the well known
result for Einstein-de Sitter model (CDM, $\Omega_{\rm{M}}=1$) for which  $t(z) = \frac{2}{3}H_0^{-1}(1 + z)^{-3/2}$. As one may conclude from the above equation, limits on the cosmological
parameters $\Omega_{\rm{M}}$ and $H_0$ (or equivalently $h$), can be derived by
fixing $t(z)$ from observations. Note also that the age
parameter, $T=H_0t_z$, depends only on the product of the two quantities $H_0$
and $t_z$, which are usually estimated from completely independent methods (Alcaniz \& Lima 1999).

McCarthy {\it et al.} sample (2004) is constituted by 20 red galaxies on the interval   $1.3<z<2.2$ selected from the Gemini Deep Deep Survey (GDDS). This sample consists of passively evolving galaxies whose most likely star formation evolution is that of a single burst of duration less than 0.1 Gyr (also consistent with 0 Gyr), thereby meaning that the galaxies evolved passively since their first star burst formation.  Another data set of interest is the Roche {\it et al} (2006) sample, which consists of 16 Extremely Red Galaxies (ERGs), formed by ellipticals, mergers and red spirals at $0.87<z<2.02$. The authors have determined a deep spectroscopic survey of these galaxies on the Great Observatories Origins Deep Survey (GOODS)-South Field by using the Gemini multi-object spectrograph on the 8-m Gemini South Telescope. They also performed an age-dating analysis by fitting the spectra and nine-band photometry ($BVIZJHK$, plus 3.6/4.5-$\mu$m fluxes from Spitzer) of the ERGS with two-component models, consisting of passively evolving, old stellar populations combined with a younger continuously star forming component. The best-fitting mean ages for the old stellar populations range from 0.6 to 4.5 Gyr with an  average value of 2.1 Gyr. The third set of data worked here is the Longhetti \etal (2007) sample, composed of nine field galaxies spectroscopically classified as early-types at $1.2<z<1.7$. They result from a near-IR spectroscopic follow-up of a complete sample of bright ERGs selected from the Munich Near-IR Cluster Survey (MUNICS) that provides optical (B, V, R, I) and near-IR (J and K') photometry. The ages of the galaxies were estimated by making use of stellar populations  with the degeneracy between the age of the best-fitting models and the star formation time scale being broken through a mass-weighting of the masses. As one may check, this sample provides the most accurate ages and the most restricting galaxies.

In Fig. 1 we  show a plot of the age as a function of the redshift for all objects present on the above quoted samples. However, since the galaxy formation is a random process it is not interesting  to consider only the ages of all objects in order to constrain the model parameters. One basic reason comes from the fact that whether a universe model is capable of explaining the existence of a very old object for a given value of $z$, it explains more easily the age of the youngest objects at that redshift. With basis on this kind of criterion we have selected a more restrictive subsample involving only the oldest objects for a fixed redshift interval. The complete subsample chosen from the 3 set of data is constituted by 13 galaxies. As can be seen in the above quoted papers, some galaxies have asymmetric age uncertainties, so we have used the D'Agostini (2004) method to symmetrize it in order to perform a $\chi^2$ statistical analysis.

In  Figs. 2 and 3 we show the selected subsample and the predictions of several $\Lambda$CDM models,  in order to verify visually the concordance with the data and the effects of the different parameters ($h$ and $\Omega_M$).

\begin{figure}[p] 
   \epsscale{1.1}
   \plotone{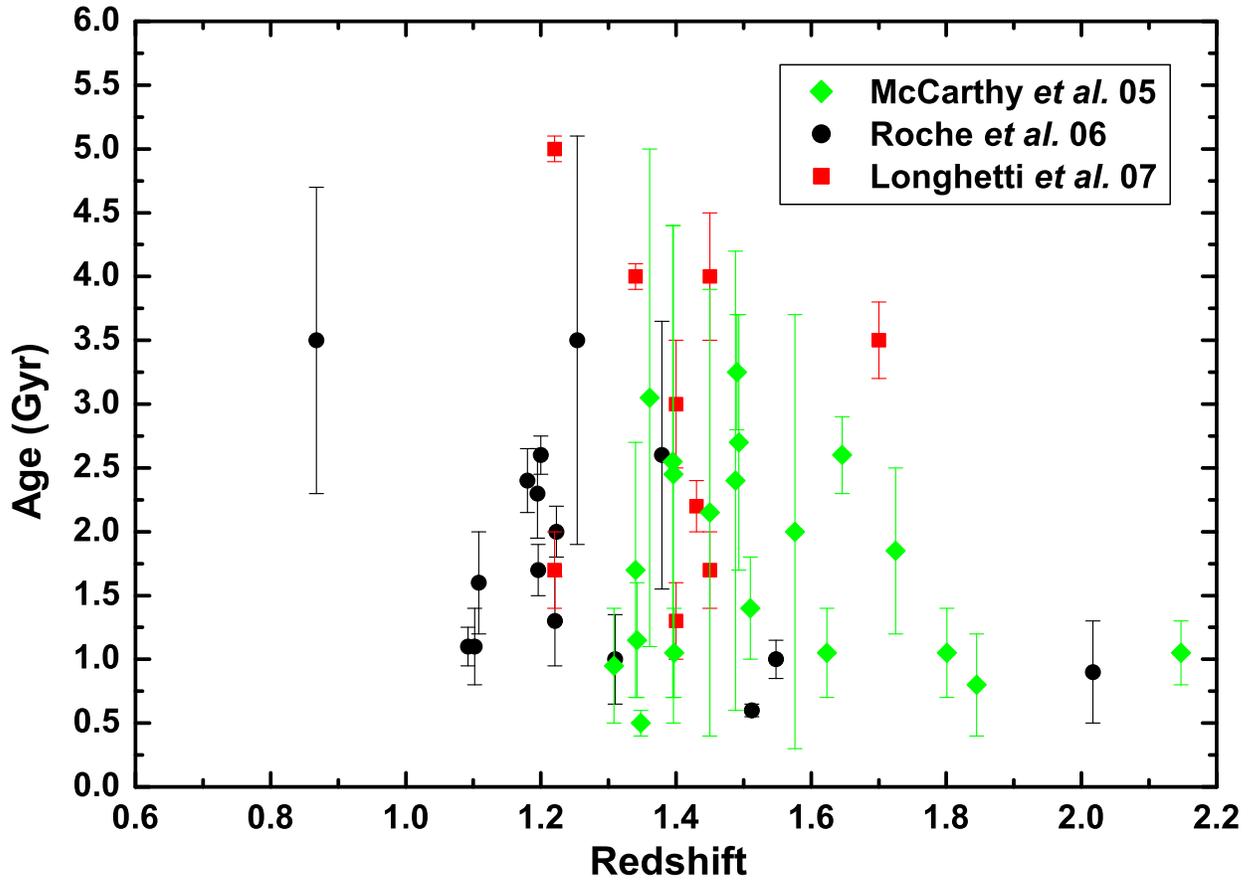}
   \caption{Plane Age-Redshift and total sample of galaxies. Green points correspond to the McCarthy {\it et al.} (2004) sample  while black and red points represent the Roche  {\it et al.} (2006) and Longhetti  {\it et al.} (2007) samples, respectively.}
\label{Fig1}
\end{figure}
\begin{figure}[p] 
   \epsscale{1.1}
   \plotone{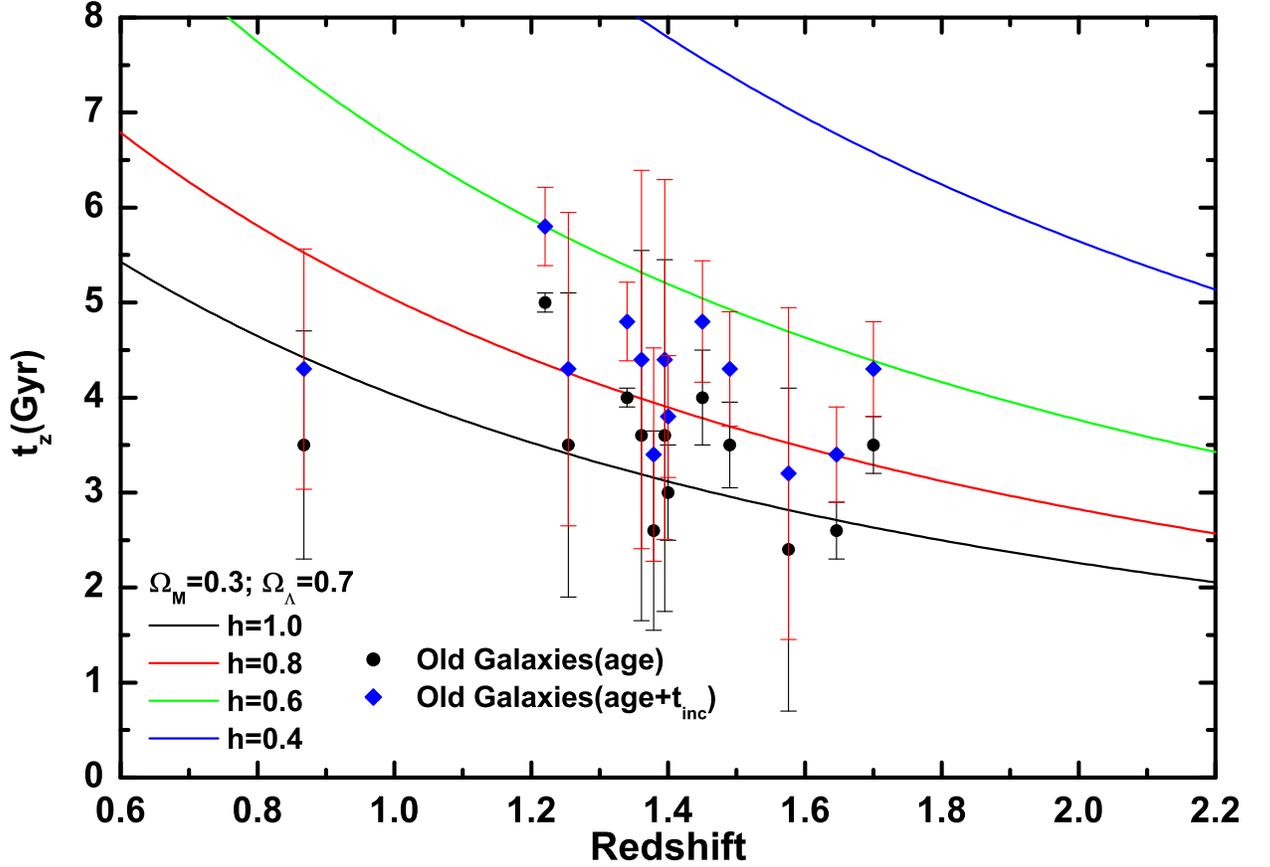}
   \caption{Age-Redshift diagram (the effect of h). The data points correspond to the 13 galaxies of our selected subsample (see the main text). The green and red filled circles stand for the observed age and the age plus the median of the incubation time ($t_{inc}=0.8Gyr$) for each object. The dotted curves are the predictions of  the cosmic  concordance model ($\Omega_M=0.3,\Omega_\Lambda=0.7$) for some different values of the $h$ parameter. As expected, for small values of $h$ it is easier for the models explain  the existence of old objects.}
\label{Fig2}
\end{figure}

\begin{figure}[p] 
   \epsscale{1.1}fig3
   \plotone{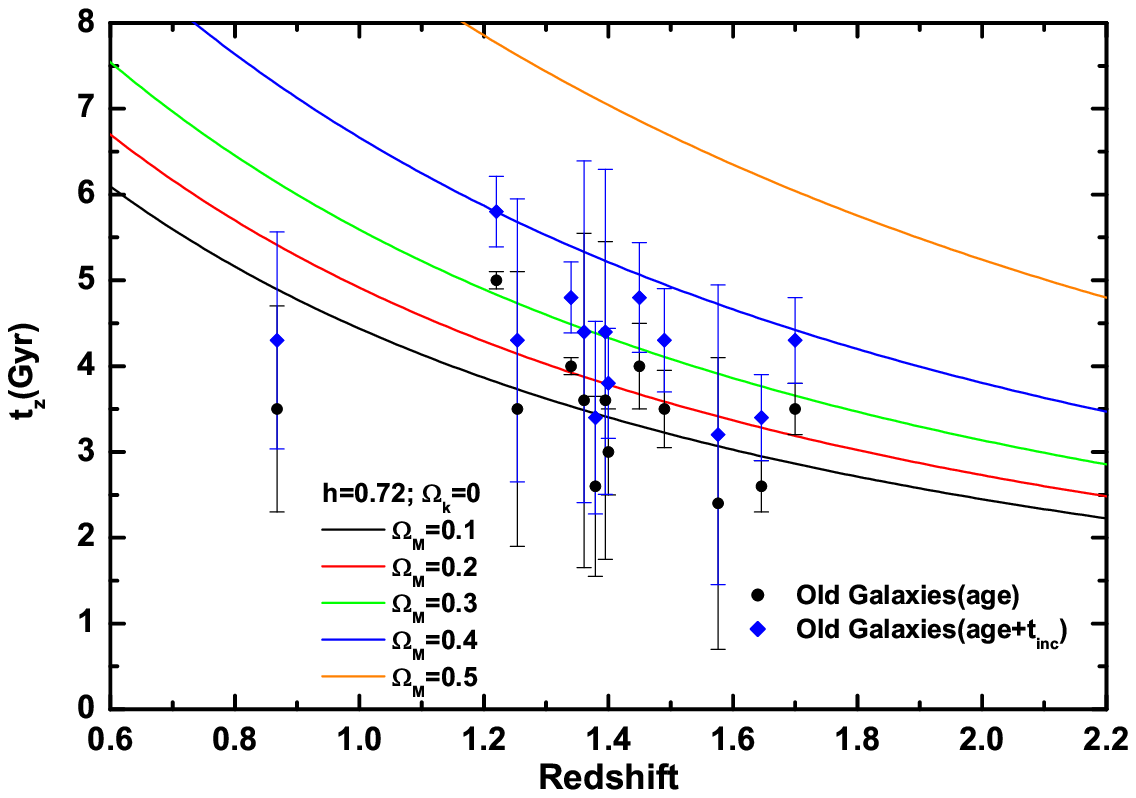}
   \caption{Age-Redshift diagram (the effect of $\Omega_M)$. As before, the data points correspond to the 13 galaxies of the selected subsample with green and red filled circles standing for the observed age and the age plus the median of the incubation time ($t_{inc}=0.8Gyr$).  Different from the previous figure, now we have plotted models with $h=0.72$, as favored by the HST Key Project, for selected values of the density parameter. For smaller values of $\Omega_M$ the ages increase thereby accommodating  the oldest selected objects.}
\label{Fig2.5}
\end{figure}


\section{Statistical Analysis and Results}

Now, let us perform a $\chi^2$ fit over the $h - \Omega_{M}$ plane.
To begin with,  we first consider that the actual  ages of the objects at a given redshift are $t_z=t_g^{obs}+t_{inc}$,  
where $t_{inc}$ is the incubation time of the galaxy. This quantity is an estimate of   the amount of time interval from the beginning of structure formation process in the Universe until the formation time ($t_f$) of the object itself. It is also assumed  that $t_{inc}$ varies slowly with the galaxy and redshift in our sample, and, in order to account for our ignorance about this kind of ``nuisance" parameter, we associate a reasonable uncertainty, say, $\sigma_{t_{inc}}$ (Fowler 1986, Sandage 1993). In what follows, we consider that  $t_{inc}=0.8 \pm 0.4$ Gyr.  Following standard lines, the maximum likelihood, ${\cal{L}} \propto \exp\left[-\chi_{age}^{2}(z;\mathbf{p}, t_{inc})/2\right]$, is determined
by a $\chi^2$ statistics
\begin{equation}
\chi^2_{age}(z|\mathbf{p}) = \sum_i { (t(z_i; \mathbf{p})-
t_{obs,i}-t_{inc})^2 \over \sigma_{t_{obs,i}}^2+\sigma_{t_{inc}}^2},
\end{equation}
where $\mathbf{p} \equiv (h, \Omega_{m})$ is the complete set of parameters, $t_{obs,i}$ is the observational age for a specific
galaxy, $\sigma_{t_{obs,i}}$ is the uncertainty in the
individual age, $\sigma_{t_{inc}}$ is the above assumed uncertainty of the incubation time. 

It now proves convenient both from a qualitative and a methodological viewpoint to consider  first the old galaxy age test
separately. Further, we present a joint analysis including the
BAO signature from the SDSS catalog. It is worth noticing that a specific flat
cosmology has not been fixed by hand in the analyzes below.

\subsection{Limits from Old Galaxies}

Let us  now consider the selected $13$ galaxies  which constitute the more stringent old galaxies present in the original samples. For the sake of definiteness this is the sample adopted for deriving all the limits in this work.  In principle,  it is reasonable to think that old galaxies at high-z alone can not put very tight constraints on the pair ($h,\Omega_M$). One of the main reasons is that the galaxy formation is a random event, and, as such, it is hard to  know  the correct values of the incubation time for each galaxy. As it appears, the incubation time for some galaxies may  be larger or smaller than the ones assumed here. 

\begin{figure*}[p] 
     \epsscale{1.1}
     \plotone{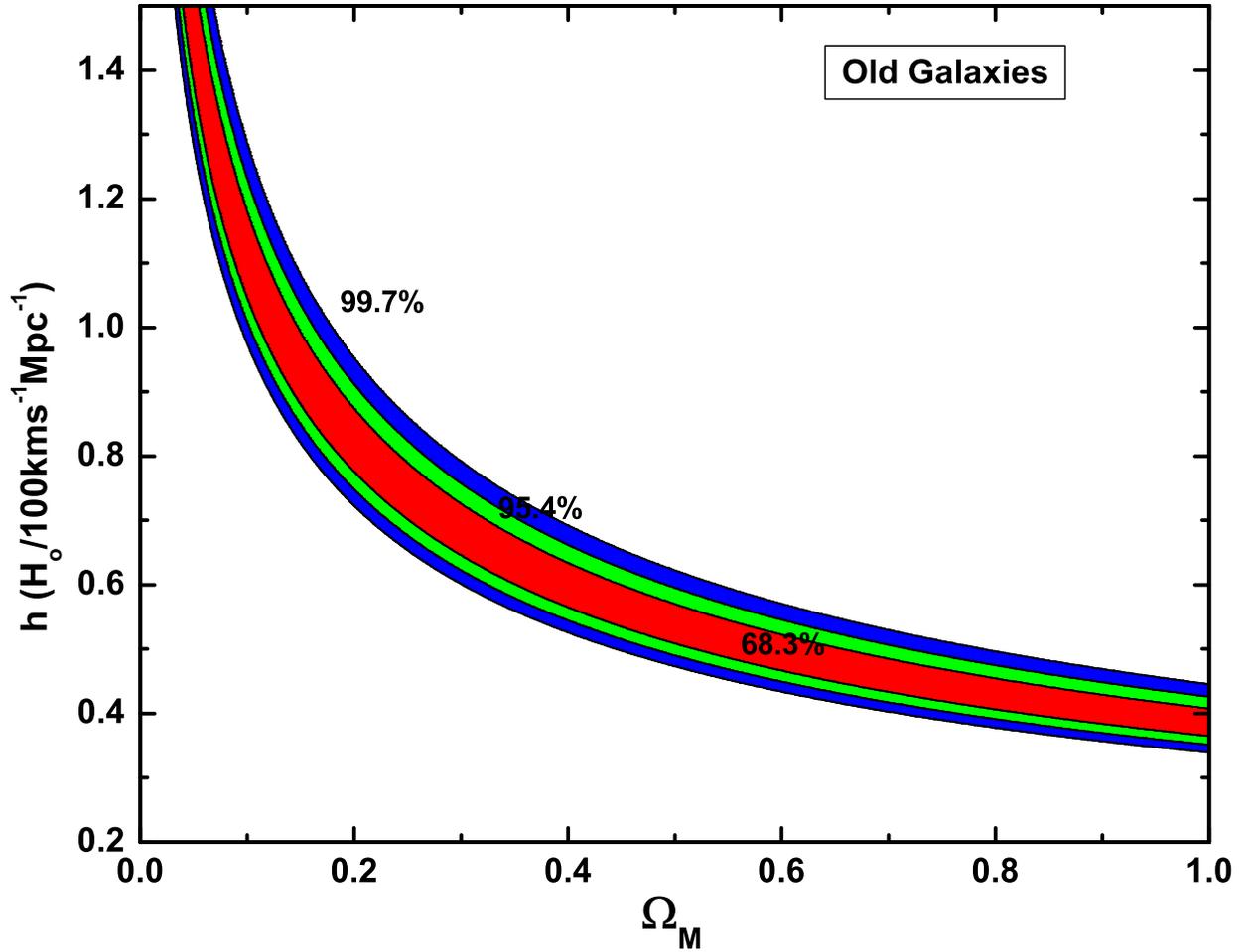}
     \caption{Confidence regions ($68.3$\%, $95.4$\% and
$99.7$\%c.l.) in the $(\Omega_{m}, h)$ plane provided by the Ages of 13 Old Galaxies at High Redshift. The best fit values are
$h = 1.11$ and $\Omega_{M} = 0.09$.} \label{Fig3}
\end{figure*}

In Fig. \ref{Fig3} we show the contours of constant likelihood (68.3\%, 95.4\% and 99.7\% c.l.) in the space parameter $h-\Omega_M$ for the
age data alone. Note that all the parameter space for $h$ and $\Omega_M$ is allowed, with a weak constraint $\Omega_M\gtrsim0.02$ and $h\gtrsim0.34$. In this case, the best fit parameters derived are $h=1.11$ and $\Omega_M=0.09$ with a $\chi^2_{min}=11.2$, however,  these values are meaningless because many others pairs of ($h,\Omega_M$) are also allowed with considerable statistical meaning. For example, by fixing $\Omega_{M}=0.3$ we have $h=0.68$ with $\chi^2_{min}=11.4$, and if $\Omega_{M}=1.0$ we have
$h=0.67$, and both cases are permitted with high degree of
confidence. Clearly, one may conclude that an additional cosmological test
(fixing $\Omega_M$) is necessary in order to break the
degeneracy on the $(h,\Omega_{M})$ plane.

\begin{figure}[p] 
     \plotone{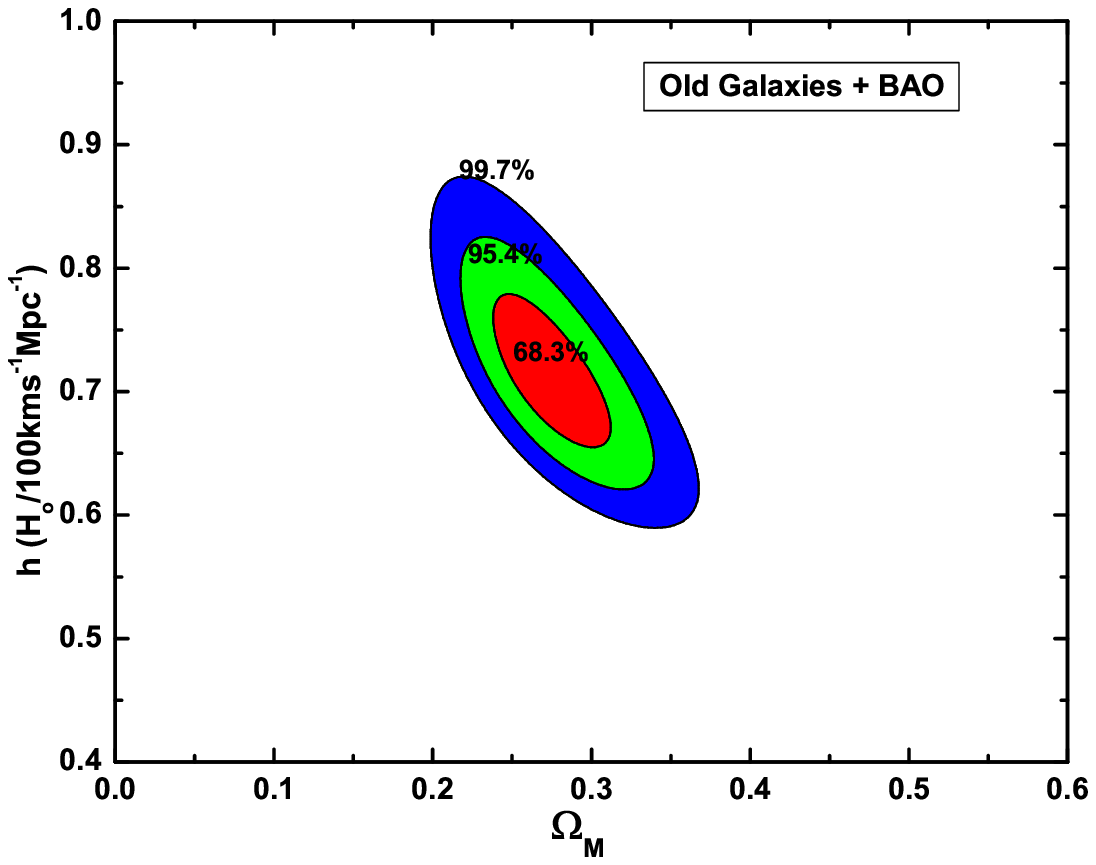}
     \caption{Contours in the $\Omega_M - h$ plane using
the Age Redshift and BAO joint analysis. The contours correspond to
$68.3$\%, $95.4$\% and $99.7$\% confidence levels. The best-fit
model converges to $h = 0.70$ and $\Omega_M=0.28$.} \label{Fig4}
\end{figure}

\subsection{Old Galaxies and BAO: A Joint Analysis}

As suggested earlier, more stringent constraints on the space
parameter ($\Omega_M$, h) can be derived by combining the
selected sample of old galaxies with the BAO signature (Eisenstein {\it{et al.}} 2005).
According to cold dark matter (CDM) picture of
structure formation, the large-scale fluctuations have grown since
$z\sim1000$ by gravitational instability. The cosmological
perturbations excited sound waves in the primeval plasma,
thereby producing the acoustic peaks in the early universe. The peak detected (from a sample of 46748 luminous red galaxies selected from the SDSS Main Sample) is predicted to arise precisely at the measured scale of 100 $h^{-1}$ Mpc. Basically, it is a consequence of the baryon acoustic oscillations in the primordial baryon-photon
plasma prior to recombination. Let us now consider it as an
additional cosmological test over the age redshift test.
As it is widely known, such a measurement can be characterized by the dimensionless parameter
\begin{eqnarray}
 {\cal{A}} \equiv {\Omega_{\rm{M}}^{1/2} \over
 {{\cal{H}}(z_{\rm{*}})}^{1/3}}\left[\frac{1}{z_{\rm{*}}}
 \Gamma(z_*)\right]^{2/3}  = 0.469 \pm 0.017, 
\label{A}
\end{eqnarray}
where $z_{\rm{*}} = 0.35$ is the redshift at which the acoustic
scale has been measured, and $\Gamma(z_*)$ is the dimensionless
comoving distance to $z_*$. The quantity ${\cal{H}}(z_{\rm{*}})$ corresponds to $H(z)/H_0$, which, in a $\Lambda$CDM flat cosmology reads:
\begin{equation}
 {\cal{H}}=[\Omega_M (1+z)^3+1-\Omega_M]^{1/2}
\end{equation}

\begin{figure*}[p] 
     \epsscale{1.1}
     \plotone{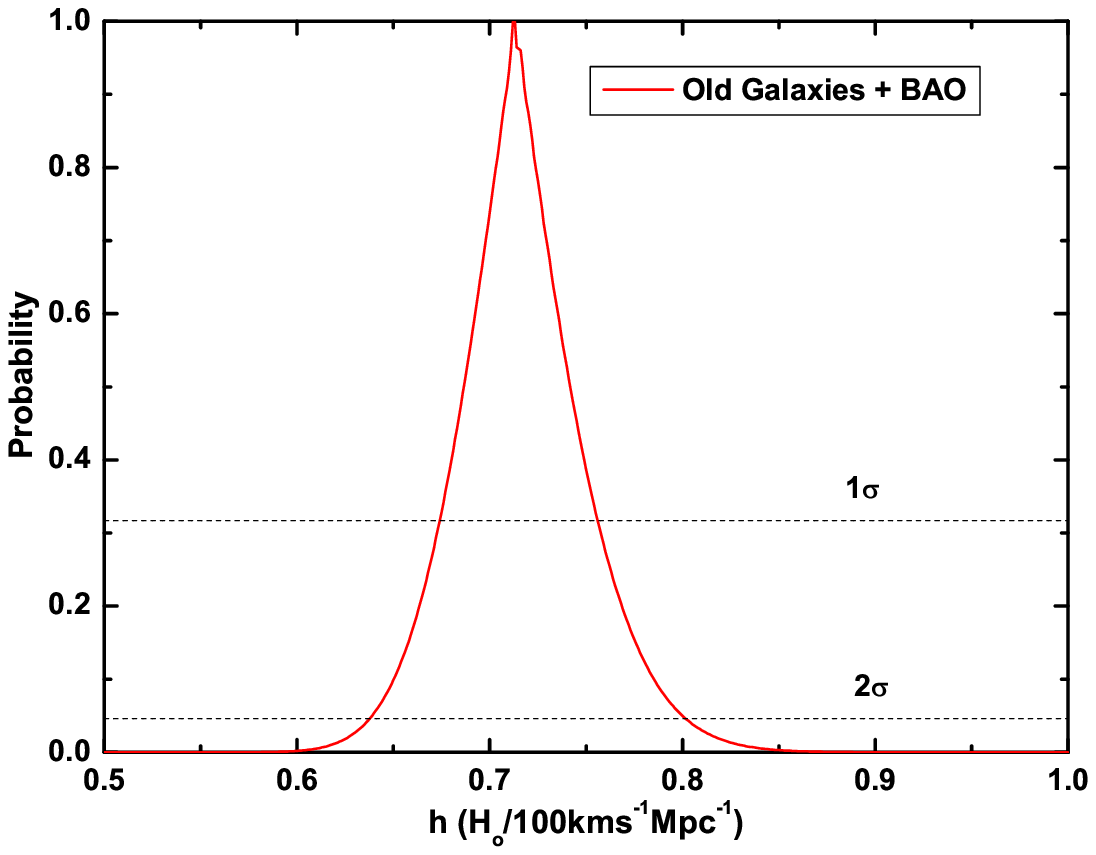}
     \caption{Likelihood function for the $h$ parameter in
a flat $\Lambda$CDM universe, from the age of old galaxies at high redshifts in combination with BAO signature. The shadow
lines are cuts in the regions of $68.3$\% and $95.4$\% probability.
We see that the region permitted is well constrained and in
concordance with others studies (Freedman {\it{et al.}} 2001;
Spergel {\it{et al.}} 2006).} \label{Fig5}
\end{figure*}
Note that the quantity given by (\ref{A}) is independent of the Hubble constant
and, as such, the BAO signature alone constrains only the $\Omega_M$
parameter. However, as can be seen from Fig. 3, the value of $h$ is heavily dependent on $\Omega_M$ thereby leading to a tighter constraint in the distance scale. This property is very characteristic of the BAO signature
thereby differentiating it from many others classical cosmological
tests, like the gas mass fraction (Lima {\it{et al.}} 2003; Allen
{\it{et al.}} 2004), luminosity distance
(Peebles \& Ratra 2003), the age of the
universe (Alcaniz {\it{et al.}} 2003; Lima 2004; Cunha \& Santos 2004; Jesus 2006), angular diameter versus redshift (Lima \& Alcaniz 2000; 2005), and the Sunyaev-Zeldovich effect (Cunha {\it et al.}  2007). 

In  Fig. \ref{Fig4}, we show the confidence regions for the 
age redshift relation and BAO joint analysis. By comparing with
Fig. \ref{Fig3}, one may see how the BAO signature breaks the
degeneracy in the $(\Omega_{\rm{M}}, h)$ plane. As it appears,
the BAO test presents a striking orthogonality centered at
$\Omega_M= 0.27 \pm {0.03}$ with respect to the age redshift data as determined from old Galaxies at High Redshift. We
find $h= 0.71\pm 0.04$ and $\chi^2_{min}=11.38$ at
$68.3$\% (c.l.) and $95.4$\% (c.l.), respectively, for $1$ free parameter. An important lesson here is that the combination of old galaxies with BAO provides one more interesting approach to constrain the Hubble constant. In Fig. \ref{Fig5}, by projecting over $\Omega_M$, we have plotted the likelihood function for the  $h$ parameter based on our joint analysis involving the BAO signature plus the OHRG data.  The dotted lines are cuts in the regions of $68.3$\% probability and $95.4$\%.

\begin{deluxetable}{lccc}
 \tablewidth{0pt} \tabletypesize{\footnotesize} \tablecaption{Limits to $H_0$\label{Tab2}}
\tablehead{{\hspace{1.5cm}} \\
\colhead{Method} & 
\colhead{Reference} & 
\colhead{h}} 
\startdata

CMB & Spergel \etal 2006 (WMAP) & $0.73\pm 0.03$  \\
Cepheid Variables & Freedman \etal 2001 (HST Project) & $0.72\pm0.08$  \\
Age Redshift & Jimenez \etal 2003 (SDSS) & $0.69\pm 0.12$  \\
SNe Ia/Cepheid & Sandage \etal 2006 & $0.62 \pm 0.013$(rand.)$\pm 0.05$(syst.)  \\
SZE+BAO & Cunha, Marassi \& Lima 2007 & $0.74^{+0.04}_{-0.03}$  \\
\textbf{Old Galaxies + BAO} & \textbf{this letter} & $0.71 \pm 0.04$  \\
\enddata
\end{deluxetable}

As can be seen from Table \ref{Tab2}, the present results are in line with some recent analyzes based on different cosmological observations, like the one provided by the HST Project (Freedman {\it{et al.}} 2001) and the WMAP team (Spergel {\it{et al.}} 2007). Note, however, that it is just marginally compatible with the determination advocated by Sandage and collaborators (2006), a result obtained with basis on Type Ia
Supernovae calibrated with Cepheid variables in nearby galaxies that hosted them. Jimenez {\it et al.} (2003) also obtained an independent estimate of the Hubble constant by computing $dz/dt$ at $z\sim0$ for a sample of galaxies from Sloan Digital Sky Survey (SDSS). By assuming a $\Lambda$CDM concordance model, they obtained $H_0=69\pm12\ksm$. More recently, Cunha, Marassi \& Lima (2007) have found $h=0.74^{+0.04}_{-0.03}$ and $\Omega_M=0.28 \pm {0.04}$ at 68\% c.l. through a joint analysis involving the  Sunyaev-Zel'dovich effect (SZe), X-ray surface brightness of galaxy clusters and baryon acoustic oscillations. They have also assumed a $\Lambda$CDM flat cosmology. The present result is in perfect agreement with their analysis with a high degree of statistical confidence (both analyzes are compatible even at 68\% c.l.). Perhaps more important, the result is heavily  dependent on the combination of the samples. For instance, if we have taken only the Roche {\it et al.} sample (the less restrictive one), after the joint analysis with BAO, the best fit increases to  $h=0.99$ ($h=0.77$ for the McCarthy {\it et al.} data).

It is also worth noticing that the best-fit scenario derived here,
$\Omega_M= 0.27 \pm {0.03}$ and  $h= 0.71\pm 0.04$,
corresponds to an accelerating Universe with $q_0=-0.58$, a total
evolutionary age of $t_o \simeq 9.6h^{-1}$ Gyr, and a transition
redshift (from deceleration to acceleration) $z_{t} \simeq 0.72$.

\section{Conclusions}

Since the original work of Hubble, the precise determination of the
distance scale, $H_0$, has been a recurrent problem in the
development of physical cosmology. In this letter we have derived
new constraints to the Hubble constant based on a statistical analysis of Ages of Old Galaxies at High Redshifts.  Our analysis revealed that with basis only on the present age-redshift data of old galaxies it is not possible to obtain tight constraints on $H_0$ due to the dependence of $t(z)$ with the $\Omega_M$ parameter (see Fig. 4). However, the degeneracy on the  $\Omega_{M}-h$ plane has been broken trough a joint analysis involving  the baryon acoustic oscillation signature from the SDSS catalog (see Fig. 5). The Hubble constant was constrained to be $h= 0.71 \pm 0.04$
($1\sigma$). These limits were derived
assuming an incubation time $t_{inc}=0.8\pm0.4$Gyr and a flat $\Lambda$CDM scenario.

As we have seen, the baryon acoustic oscillations (BAO)  provide an interesting
tool for constraining directly the mass density parameter,
$\Omega_M$, and, indirectly, it also improves the Hubble constant
limits acquired from other cosmological techniques (like the
age-redshift relation). Our Hubble constant estimation through a joint 
analysis  involving old high redshift galaxies plus BAO is in good agreement with independent estimates  based on different cosmological observations, like the third year of the WMAP and the HST Key Project. Indirectly, such an agreement suggests that
the incubation time we have assumed is quite realistic. It also
reinforces the interest to the observational search of old galaxies at high redshifts 
in the near future, when more and larger samples, smaller
statistical and systematic uncertainties can improve the limits on
the present value of the Hubble parameter.

\section*{Acknowledgments}
The authors are grateful to L. Marassi and R. C. Santos for helpful discussions. JASL is partially supported by CNPq and FAPESP (No. 04/13668-0). JFJ is supported by CNPq and JVC by
FAPESP (No. 05/02809-5). 

{}

\end{document}